\documentstyle{article}
\title{An axiomatic framework for classical particle mechanics without space-time\thanks{This paper has been accepted for publication in {\em Philosophia Naturalis.\/}}}
\author{Adonai S. Sant'Anna\\Dep. Matem\'atica, Universidade Federal do Paran\'a\\C.P. 19081, Curitiba, PR, 81531-990, Brazil\and Phone:+55-41-246-3518 (Voice and FAX)\\e-mail:adonai@mat.ufpr.br}
\date{ }
\begin{document}
\maketitle

\begin{abstract}
We present an axiomatic framework for non-relativistic classical particle mechanics, inspired on Tati's ideas about a non-space-time description for physics. The main advantage of our picture is that it allows us to describe causality without any reference to elapsed time intervals.
\end{abstract}

\newtheorem{definicao}{Definition}[section]
\newtheorem{teorema}{Theorem}[section]
\newtheorem{lema}{Lemma}[section]
\newtheorem{corolario}{Corolary}[section]
\newtheorem{proposicao}{Proposition}[section]
\newtheorem{axioma}{Axiom}[section]

\section{Introduction}

	Before the famous experiment by Michelson and Morley in 1887, physicists believed that there should exist an ether in space, in order to explain the propagation of electromagnetic waves, by means of the mechanical theory of waves. However, that experiment showed that there is no relative motion of our planet with respect to a physical medium usually refered to as the ether.

	Now there is the belief that there should exist space-time as a medium which allows to order physical events. The `new' ether is a collection of physical properties of a continuum space-time (Einstein, 1991). In this paper we propose a discrete picture for the dynamics in classical particle mechanics, where the continuum has no physical meaning at all. We work on the possibility that the world is, in some sense, atomistic. And space-time, as one of the constituents of the world, is also atomistic. We show that we may have causality without any time interval, in the usual sense, between two events.

	Some decades ago the japanese physicist T. Tati began a researche program about a description for physical theories where the concepts of space and time are not {\em primitive\/} notions (Tati, 1964, 1983, 1986, 1987). For a quick reference on his work see (Tati, 1986). For a detailed presentation of the non-space-time picture of classical mechanics, electromagnetism, quantum mechanics and quantum electrodynamics (QED) see (Tati, 1964). Tati's objective is to solve a specific problem, namely, the divergences in quantum field theory. Tati argues that it is possible {\em to define} space and time in some physical theories, by using the description proposed by him. Space and time are fundamental concepts that should remain in classical physics. But in quantum field theories, the classical approach is meaningless. So, space and time do not exist at microscopic levels. Tati's work supports a theory of finite degree of freedom in QED, which allows to eliminate the divergences.

	Although Tati's motivation was a {\em physical\/} problem, we consider that a non-space-time description for physical theories is also interesting from the {\em philosophical\/} point of view. The investigation of all logically possible physical theories may conduct to a better understanding of the real role of fundamental concepts and principles in physical theories. For example, we have recently showed how to define a set-theoretical predicate (in the sense of P. Suppes' (1967) program of axiomatization) for classical particle mechanics without the concept of force (Sant'Anna, 1995; 1996), based on Hertz's mechanics (Hertz, 1894; 1956). Now, we are defining an axiomatic framework for non-relativistic classical particles mechanics without space-time.

	Thus, our work is in agreement with P. Suppes' words about the role for philosophy in the sciences:

\begin{quote}
We are no longer Sunday's preachers for Monday's scientific workers, but we can participate in the scientific enterprise in a variety of constructive ways. Certain foundational problems will be solved better by philosophers than by anyone else. Other problems of great conceptual interest will really depend for their solution upon scientists deeply immersed in the discipline itself, but illumination of the conceptual significance of the solutions can be a proper philosophical role. (Suppes, 1990).
\end{quote}

	Tati does not assume space-time as a primitive notion. Rather than space-time, he considers {\em causality\/} as a primitive concept, whatever it does really mean.

	We do not present in this paper all details of Tati's theory because we consider his formulation a bit confuse from the logico-mathematical standpoint. So, our starting point is the intuition presented in Tati's work which, we believe, is somehow preserved in our paper (at least in principle).

	We venture to interpret Tati's work at our own risk, as it follows. Physical observations may be associated to elements of a discrete set which corresponds, intuitivelly speaking, to {\em state\/} measurement values. Each observation is related to another one by a causal relation. This causal relation may be expressed by equations which are very similar to numerical solutions of differential equations commonly used in physics. The notion of a {\em continuum\/} space-time in theoretical physics allows the use of standard differential and integral calculus on those equations. Space-time intervals, which are associated to elements of a {\em continuum}, should be regarded as `unknowables', if we use Tati's terminology. Such unknowables behave like hidden variables in the sense that we cannot actually measure {\em real\/} space-time intervals. Measurements do not have arbitrary precision. It should also be emphasized that even measurements of mass, position, momentum, etc., should be associated to elements of a discrete set of numbers, if we are interested to eliminate anthropomorphical notions like the {\em continuum} and real numbers.

	We recall the well known words by L. Kronecker: ``God made the integers, all the rest is the work of man.''

	In the next Section we present the axiomatic framework for classical particle mechanics by McKinsey, Sugar and Suppes. In Section 3, a non-space-time description for classical particle mechanics is presented, based on the formulation described in Section 2. In Section 4 we show that our description is consistent. In Section 5 we discuss some possible applications of our picture, as well as other related lines for future works.

\section{McKinsey-Sugar-Suppes Predicate for Classical Particle Mechanics}

	This section is essentially based on the axiomatization for classical particle mechanics presented in (Suppes, 1957), which is a variant of the formulation in (McKinsey et al., 1953). We abbreviate McKinsey-Sugar-Suppes System for Classical Particle Mechanics as M.S.S. system. Our intention is to apply Tati's ideas on the M.S.S. system in order to illustrate our non-space-time description of physics. We have chosen M.S.S. system because it represents the simplest case of a physical theory ever axiomatized.

	M.S.S. system has six primitive notions: $P$, $T$, $m$, $s$, $f$, and $g$. $P$ and $T$ are sets; $m$ is a vector-valued unary function defined on $P$; $s$ and $g$ are vector-valued binary functions defined on the Cartesian product $P\times T$; and $f$ is a vector-valued ternary function defined on the Cartesian product $P\times P\times T$. Intuitivelly, $P$ corresponds to the set of particles and $T$ is to be physically interpreted as a set of real numbers measuring elapsed times (in terms of some unit of time, and measured from some origin of time). $m(p)$ is to be interpreted as the numerical value of the mass of $p\in P$. $s_{p}(t)$, where $t\in T$, is a $3$-dimensional vector which is supposed to be physically interpreted as the position of particle $p$ at instant $t$. $f(p,q,t)$, where $p$, $q\in P$, corresponds to the internal force that particle $q$ exerts over $p$, at instant $t$. Finally, the function $g(p,t)$ is to be understood as the external force acting on particle $p$ at instant $t$.

	Now, we can give the axioms for M.S.S. system.

\begin{definicao}
${\cal P} = \langle P,T,s,m,f,g\rangle$ is a M.S.S. system if and only if the following axioms are satisfied:

\begin{description}
\item [P1] $P$ is a non-empty, finite set.
\item [P2] $T$ is an interval of real numbers.
\item [P3] If $p\in P$ and $t\in T$, then $s_{p}(t)$ is a
$3$-dimensional vector such that $\frac{d^{2}s_{p}(t)}{dt^{2}}$ exists.
\item [P4] If $p\in P$, then $m(p)$ is a positive real number.
\item [P5] If $p,q\in P$ and $t\in T$, then $f(p,q,t) = -f(q,p,t)$.
\item [P6] If $p,q\in P$ and $t\in T$, then $[s_{p}(t), f(p,q,t)] =
-[s_{q}(t), f(q,p,t)]$.
\item [P7] If $p,q\in P$ and $t\in T$, then
$m(p)\frac{d^{2}s_{p}(t)}{dt^{2}} = \sum_{q\in P}f(p,q,t) + g(p,t).$
\end{description}
\end{definicao}

	The brackets [,] in axiom {\bf P6} denote vector product.

	Axiom {\bf P5} corresponds to a weak version of Newton's Third Law: corresponding to every force there is always a counterforce. Axioms {\bf P6} and {\bf P5}, correspond to the strong version of Newton's Third Law, since axiom {\bf P6} establishes that the direction of force and counterforce is the direction of the line between particles $p$ and $q$.

	Axiom {\bf P7} corresponds to Newton's Second Law.

\begin{definicao}
Let ${\cal P} = \langle P,T,s,m,f,g\rangle$ be a M.S.S. system, let $P'$ be a non-empty subset of $P$, let $s'$, $g'$, and $m'$ be the functions $s$, $g$, and $m$ with their first arguments restricted to $P'$, and let $f'$ be the function $f$ with its domain $P\times P\times T$ restricted to $P'\times P'\times T$. Then ${\cal P'} = \langle P',T,s',m',f',g'\rangle$ is a subsystem of ${\cal P}$ if the following condition is satisfied:

\begin{equation}\label{P7p}
m'(p)\frac{d^{2}s'_{p}(t)}{dt^{2}} = \sum_{q\in P'}f'(p,q,t) + g'(p,t).
\end{equation}
\label{P7}
\end{definicao}

	Actually, in (Suppes, 1957), definition (\ref{P7}) does not have equation (\ref{P7p}). On the other hand, such an equation is really necessary to prove the following theorem:

\begin{teorema}
Every subsystem of a M.S.S. system is again a M.S.S. system.
\end{teorema}

\begin{definicao}
Two M.S.S. systems \[{\cal P} = \langle P,T,s,m,f,g\rangle\] and \[{\cal P'} = \langle P',T',s',m',f',g'\rangle\] are equivalent if and only if $P=P'$, $T=T'$, $s=s'$, and $m=m'$.
\end{definicao}

\begin{definicao}
A M.S.S. system is isolated if and only if for every $p\in P$ and $t\in T$, $g(p,t) = {\bf 0}$, where ${\bf 0}$ is the null vector.
\end{definicao}

\begin{teorema}
If \[{\cal P} = \langle P,T,s,m,f,g\rangle\] and \[{\cal P'} = \langle P',T',s',m',f',g'\rangle\] are two equivalent systems of particle mechanics, then for every $p\in P$ and $t\in T$
\[\sum_{q\in P}f(p,q,t)+g(p,t) = \sum_{q\in P'}f'(p,q,t)+g'(p,t).\]\label{somaforcas}
\end{teorema}

	The imbedding theorem is the following:

\begin{teorema}
Every M.S.S. system is equivalent to a subsystem of an isolated system of particle mechanics.\label{Her}
\end{teorema}

\section{Classical Particle Mechanics Without Space-Time}

	In this section we define a set-theoretical predicate for a classical particle mechanics system, inspired on Tati's ideas. By set-theoretical predicate we mean Suppes predicate (Suppes-1967). We are aware about the limitations of particle mechanics, but that is not the point. The issue is Tati's picture for physical theories. Obviously we intend to apply these same ideas on other physical theories. But that is a task for future papers. Our main goal in the present work is to give an axiomatic framework for a simple case of physical theory, where space-time is not stated as one of the primitive notions. The simplest case of a physical theory, in our opinion, is M.S.S. system.

	In this paragraph we settle some notational features to be used in the paper from now on. We denote the set of real numbers by $\Re$, the set of integer numbers by ${\bf Z}$, the set of positive integers by ${\bf Z}^+$ and the cartesian products $\Re\times\Re\times\Re$ and ${\bf Z}\times{\bf Z}\times{\bf Z}$, respectively by $\Re^3$ and ${\bf Z}^3$. When there is no risk of confusion, we say that $\Re^3$ is a real vector space. We say that $h$ is a $(C^0,\tau)$-function iff $h = h(\tau)$ and $h$ is continuous with respect to $\tau$. Moreover, we say that $h$ is a $(C^k,\tau)$-function iff $h = h(\tau)$ and it is continuous and continuously differentiable (with respect to $\tau$) $k$ times.

	Our system has sixteen primitive concepts: $P$, $I$, $T$, $m$, $\bar{s}$, $\bar{f}$, $\bar{g}$, $s$, $v$, $f$, $g$, $c_s$, $c_v$, $c_f$, $c_g$, and $c_t$.

\begin{definicao}
${\cal F} = \langle P, I, T, m, \bar{s}, \bar{f}, \bar{g}, s, v, f, g, c_s, c_v, c_f, c_g, c_t\rangle$ is a {\em non-relativistic classical particle mechanics system without space-time}, which we abbreviate as CM-Tati's system, if and only if the following axioms are satisfied:

\begin{description}

\item[CM-1] $P$ is a non-empty finite set;

\item[CM-2] $I\subset {\bf Z^+}$;

\item[CM-3] $T$ is an interval of real numbers;

\item[CM-4] $m:P\to{\bf Z}^+$ is a function whose images are denoted by $m^p$;

\item[CM-5] $s:P\times I\to{\bf Z}^3$ is a function whose images are denoted by $s_i^p$, where $i\in I$ and $p\in P$. Yet, if $p\neq p'$ then $s_i^p\neq s_i^{p'}$;

\item[CM-6] $v:P\times I\to{\bf Z}^3$ is a function whose images are denoted by $v_i^p$, where $i\in I$ and $p\in P$;

\item[CM-7] $f:P\times P\times I\to{\bf Z}^3$ is a function whose images are denoted by $f_i^{pq}$, where $i\in I$ and $p, q \in P$;

\item[CM-8] $g:P\times I\to{\bf Z}^3$ is a function whose images are denoted by $g_i^p$, where $i\in I$ and $p\in P$;

\item[CM-9] $c_t:I\to T$ is a function whose images are denoted by $c_t(i)$, where $i\in I$;

\item[CM-10] $\bar{s}:P\times T\to\Re^3$ is a $(C^2,\tau)$-function whose images are denoted by $\bar{s}^p(\tau)$, satisfying the following property: if there exists $i\in I$ such that $\tau = c_t(i)$, then $\bar{s}^p(\tau) = s_i^p$;

\item[CM-11] If there exists $i\in I$ such that $\tau = c_t(i)$, then $\frac{d}{d\tau}\bar{s}^p(\tau) = v_i^p$;

\item[CM-12] $\bar{f}:P\times P\times T\to\Re^3$ is a $(C^0,\tau)$-function whose images are denoted by $\bar{f}^{pq}(\tau)$, satisfying the following property: if there exists $i\in I$ such that $\tau = c_t(i)$, then $\bar{f}^{pq}(\tau) = f_i^{pq}$;

\item[CM-13] $\bar{g}:P\times T\to\Re^3$ is a $(C^0,\tau)$-funcion whose images are denoted by $\bar{g}^p(\tau)$, satisfying the following property: if there exists $i\in I$ such that $\tau = c_t(i)$, then $\bar{g}^p(\tau) = g_i^p$;

\item[CM-14] For all $p,q\in P$ and $\tau\in T$ we have $\bar{f}^{pq}(\tau) = -\bar{f}^{qp}(\tau)$;

\item[CM-15] For all $p,q\in P$ and $\tau\in T$ we have $[\bar{s}^p(\tau), \bar{f}^{pq}(\tau)] = -[\bar{s}^p(\tau), \bar{f}^{qp}(\tau)]$;

\item[CM-16] For all $p,q\in P$ and $\tau\in T$ we have \[m^p\frac{d^{2}\bar{s}^p(\tau)}{d\tau^2} = \sum_{q\in P}\bar{f}^{pq}(\tau) + \bar{g}^p(\tau);\]

\item[CM-17] $c_s$ is a recursive function such that $s_{i+1}^p = c_s(s_i^p,v_i^p,f_i^{pq},g_i^p)$;

\item[CM-18] $c_v$ is a recursive function such that $v_{i+1}^p = c_v(s_i^p,v_i^p,f_i^{pq},g_i^p)$;

\item[CM-19] $c_f$ is a recursive function such that $f_{i+1}^{pq} = c_f(s_i^p,v_i^p,f_i^{pq},g_i^p)$;

\item[CM-20] $c_g$ is a recursive function such that $g_{i+1}^p = c_g(s_i^p,v_i^p,f_i^{pq},g_i^p)$;

\item[CM-21] The diagram

\[P\times P\times I\stackrel{\mbox{$\gamma$}}{\longrightarrow}{\bf Z}^3\times{\bf Z}^3\times{\bf Z}^3\times{\bf Z}^3\]
\[\varphi\downarrow\;\;\;\;\;\;\;\;\;\;\;\;\;\;\;\;\;\;\;\;\;\;\;\;\;\;\downarrow c\;\;\;\]
\[P\times P\times I\stackrel{\mbox{$\gamma$}}{\longrightarrow}{\bf Z}^3\times{\bf Z}^3\times{\bf Z}^3\times{\bf Z}^3\]

\noindent
commutes, where $\varphi(p,q,i) = (p,q,i+1)$, $\gamma(p,q,i) = (s_i^p,v_i^p,f_i^{pq},g_i^p)$ and $c(s_i^p,v_i^p,f_i^{pq},g_i^p) =$\\ $(c_s(s_i^p,v_i^p,f_i^{pq},g_i^p),c_v(s_i^p,v_i^p,f_i^{pq},g_i^p),c_f(s_i^p,v_i^p,f_i^{pq},g_i^p),c_g(s_i^p,v_i^p,f_i^{pq},g_i^p))$;

\end{description}

\end{definicao}

	This paragraph follows with some aditional intuitive hints about our axioms. $P$ corresponds to the set of particles. Axiom {\bf CM-1} says that we are dealing only with a finite number of particles. Axiom {\bf CM-2} says that $I$ is a set of positive integers; intuitivelly, each $i\in I$ corresponds to an {\em observation\/}. When we refer to observations we  talk about either {\em performed\/} observations or {\em potentially performable\/}(in a sense) observations. Axiom {\bf CM-3} says that $T$ is an interval of real numbers. The intuitive meaning of $T$ is clarified when we discuss about the function $c_t$. In M.S.S. system, $T$ is interpreted as time. We say that $m^p$, in axiom {\bf CM-4}, corresponds to the (inertial) mass of particle $p$, which is a positive integer number. Such a condition demands an adequate measurement unit for mass, obviously different from the usual units. $s_i^p$ in axiom {\bf CM-5} corresponds to the position of particle $p$ at the $i$-th observation, while $v_i^p$ in axiom {\bf CM-6} is the speed of particle $p$ at observation $i$. In axiom {\bf CM-7} $f_i^{pq}$ corresponds to the internal force that particle $q$ exerts over $p$, at the $i$-th observation. In the next axiom $g_i^p$ is interpreted as the external force over $p$ at observation $i$. Function $c_t$ in axiom {\bf CM-9} is the correspondence that physicists make between their observations and the working of an ideal (in some sense) chronometer, represented by $T$. It should be emphasized that we are not imposing that $c_t$ is an injective function. That means that we may have causal relations without any passage of a `time' interval $[0,\tau]$, which is a very common situation in quantum mechanics as well as in the usual descriptions for classical mechanics. See, for instance, the problem of instantaneous actions-at-a-distance in newtonian mechanics. Functions $\bar{s}$, $\frac{d}{d\tau}\bar{s}$, $\bar{f}$ and $\bar{g}$ in axioms {\bf CM-10}, {\bf CM-11}, {\bf CM-12} and {\bf CM-13} are extensions of, respectively, $s$, $v$, $f$ and $g$. Such extensions allow us to use differential and integral calculus according to axiom {\bf CM-16}, although there is no physical interpretation for $\bar{s}$, $\frac{d}{d\tau}\bar{s}$, $\bar{f}$, and $\bar{g}$ when there is no $i\in I$ such that $\tau = c_t(i)$. Axioms {\bf CM-14}, {\bf CM-15} and {\bf CM-16} are analogous to axioms {\bf P5}, {\bf P6} and {\bf P7} in M.S.S. system. Functions $c_s$, $c_v$, $c_f$ and $c_g$ are called `causal relations'. The commutative diagram in axiom {\bf CM-21} is necessary in order to state a sort of compatibility between the causal relations and positions, speeds and forces. For those who are not familiar with the language of category theory, we say that the diagram in axiom {\bf CM-21} commutes when $\gamma\circ c = \varphi\circ\gamma$, where $\circ$ referes to functions composition.

	In the following theorems we denote the imbedding function by $1$.

\begin{teorema}\label{1}
The diagram

\[P\times I\stackrel{\mbox{\it s}}{\longrightarrow}{\bf Z}^3\]
\[\;\;\;\alpha\downarrow\;\;\;\;\;\;\;\;\;\;\;\;\downarrow 1\]
\[P\times T\stackrel{\bar{\mbox{\it s}}}{\longrightarrow}\Re^3\]

\noindent
commutes, where $\alpha(p,i) = (p,c_t(i))$.
\end{teorema}

\noindent
{\bf Proof:} direct from axiom {\bf CM-10}.

\begin{teorema}\label{2}
The diagram

\[P\times I\stackrel{\mbox{\it v}}{\longrightarrow}{\bf Z}^3\]
\[\;\;\;\alpha\downarrow\;\;\;\;\;\;\;\;\;\;\;\;\downarrow 1\]
\[P\times T\stackrel{\frac{d}{d\tau}\bar{\mbox{\it s}}}{\longrightarrow}\Re^3\]

\noindent
commutes, where $\alpha(p,i) = (p,c_t(i))$.
\end{teorema}

\noindent
{\bf Proof:} direct from axiom {\bf CM-11}.

\begin{teorema}\label{3}
The diagram

\[P\times P\times I\stackrel{\mbox{\it f}}{\longrightarrow}{\bf Z}^3\]
\[\;\;\;\;\;\;\beta\downarrow\;\;\;\;\;\;\;\;\;\;\;\;\;\;\;\;\downarrow 1\]
\[P\times P\times T\stackrel{\bar{\mbox{\it f}}}{\longrightarrow}\Re^3\]

\noindent
commutes, where $\beta(p,q,i) = (p,q,c_t(i))$.
\end{teorema}

\noindent
{\bf Proof:} direct from axiom {\bf CM-12}.

\begin{teorema}\label{4}
The diagram

\[P\times I\stackrel{\mbox{\it g}}{\longrightarrow}{\bf Z}^3\]
\[\;\;\;\alpha\downarrow\;\;\;\;\;\;\;\;\;\;\;\;\downarrow 1\]
\[P\times T\stackrel{\bar{\mbox{\it g}}}{\longrightarrow}\Re^3\]

\noindent
commutes, where $\alpha(p,i) = (p,c_t(i))$.
\end{teorema}

\noindent
{\bf Proof:} direct from axiom {\bf CM-13}.\\

	The commuting diagram in axiom {\bf CM-21} describes the dynamics of our discrete picture for physical phenomena. Obviously, such a dynamics does not depend on time $T$. Nevertheless, the diagrams of theorems (\ref{1}), (\ref{2}), (\ref{3}) and (\ref{4}) show the compatibility of the discrete picture and the continuous description given in axioms {\bf CM-10} $\sim$ {\bf CM-16}, which are in correspondence, in a certain sense, with some of the axioms of M.S.S. system. Those diagrams suggest that the natural language for CM-Tati's system is category theory. It seems reasonable to describe the causal relations $c_s$, $c_v$, $c_f$ and $c_g$ as morphisms of a given category, whose objects are in correspondence with the observations of $s_i^p$, $v_i^p$, $f_i^{pq}$ and $g_i^p$.

	The definitions of subsystem, isolated system and equivalent systems, as well as the corresponding theorems (see Section 2), can be easily stated.

\section{CM-Tati's System is Consistent}

	In this section we present a model for CM-Tati's system. Axioms {\bf CM-1}, {\bf CM-3} and {\bf CM-10}$\sim${\bf CM-16} are coincident with M.S.S. system assumptions up to function $c_t$ and to the fact that mass is a positive integer. There is no trouble with respect to integer mass. We may assume M.S.S. system with integer masses as our model. We also consider that $c_t$ is the imbedding function, if we make an appropriate choice for $I$ and $T$ ($I = \{1,2,3,4\}$ and $T = [1,4]$, for example). Now, the question is: how to interpret functions $s$, $v$, $f$, $g$, $c_s$, $c_v$, $c_f$, and $c_g$ and the diagram given in axiom {\bf CM-21}.

	Let us consider $f$ and $g$ as constants with respect to $i$, i.e., $\forall i \forall i' ((i\in I, i'\in I) \to (f_i^{pq} = f_{i'}^{pq} \wedge g_i^p = g_{i'}^p))$, where $\wedge$ and $\forall$ stand, respectivelly, for the logical connective `and' and the universal quantifier. Consider also that $\bar{f}$ and $\bar{g}$ are constants with respect to $\tau$ and that $\forall i( i\in I \to c_f(s_i^p,v_i^p,f_i^{pq},g_i^p) = f_i^{pq} \wedge c_g(s_i^p,v_i^p,f_i^{pq},g_i^p) = g_i^p)$, i.e., $f_{i+1}^{pq} = f_i^{pq}$ and $g_{i+1}^p = g_i^p$.

	Under these assumptions, the solution of the differential equation given in axiom {\bf CM-16} is:

\begin{equation}
\bar{s}^p(\tau) = \bar{s}^p(\tau_0) + \frac{d\bar{s}^p(\tau_0)}{d\tau}(\tau-\tau_0) + \frac{1}{2}\frac{\sum_{q\in P}\bar{f}^{pq}(\tau_0) + \bar{g}^p(\tau_0)}{m^p}(\tau-\tau_0)^2.\label{horas}
\end{equation}

	This solution is valid, in particular, for integer values $\tau_0 = i$ and $\tau = i+1$ according to our statement that $c_t$ is the imbedding function. In this case it is easy to verify the commutativity of the diagram in axiom {\bf CM-21} if we consider that $s_{i+1}^p = c_s(s_i^p,v_i^p,f_i^{pq},g_i^p)$ is equivalent to equation (\ref{horas}) when $\tau = i+1$ and $\tau_0 = i$. And adequate choice for the values of forces $f$ and $g$, as well as to the mass $m$, will conduct to integer solutions $s$ of equation \ref{horas}. That means that we should choose adequate measurement units for mass, force, speed and position.

\section{Final Remarks}

	There are some open problems related to our discrete picture for classical particle mechanics:

\begin{enumerate}

\item How to describe transformations of coordinates systems in ${\bf Z}^3$? What is a coordinate system in ${\bf Z}^3$? Concepts like invariance and covariance should be revised. We know that ${\bf Z}^3$ may be taken as a free ${\bf Z}$-module, where ${\bf Z}$ is a ring, i.e., ${\bf Z}^3$ may have a basis and a respective dimension. We may also define rotations and translations with respect to a given basis. Nevertheless, a group of transformations in this discrete space is a task for future works.

\item How to extend such ideas to continuum mechanics? Is there any continuum mechanics in this picture?

\item How to extend our picture to quantum mechanics or quantum field theory? That is an interesting point, since in EPR (Einstein-Podolsky-Rosen experiment) we have an instantaneous and non-local interaction between two elementary particles. If we extend our picture to quantum physics we may be able to explain EPR as a causal relation with no elapsed time interval, by considering that $c_t$ is not injective. That does not violate special relativity if we consider that relativity holds only when $c_t$ is injective.

\item How to make a discrete picture for physical theories, without any reference to continuum spaces? It is well known that Turing machines do not provide an adequate notion for computability in the set of real numbers. If a discrete picture for physics is possible, some undecidable problems which are very common in theoretical physics should disappear.

\end{enumerate}

	These questions cannot be answered in this paper. But we certainly intend to do it in future works.

\section{Acknowledgments}

	We gratefully acknowledge the important criticisms and suggestions made by D\'ecio Krause (Universidade Federal do Paran\'a, Brazil), Heinz-J\"urgen Schmidt (Universit\"at Osnabr\"uck, Germany), and Newton da Costa (Universidade de S\~ao Paulo, Brazil). We were also benefitted by criticisms of one anonymous referee of this journal.\\\\\\
\noindent
{\em References}
\small
\begin{enumerate}
\item Einstein, A., 1991, `On the ether', in S. Saunders and H.R. Brown (eds.) {\em The Philosophy of Vacuum}, Oxford Un. Press, New York, 13-20. Originally published as `\"Uber den \"Ather', 1924, {\em Schweizerische naturforschende Gesellschaft, Verhanflungen\/}, {\bf 105}, 85-93.
\item Hertz, H.R., 1894, {\em Die Prinzipien der Mechanik in Neuem Zusammenhange Dargestellt}, Barth, Leipzig.
\item Hertz, H.R., 1956, {\em The Principles of Mechanics}, English translation by D.E. Jones and J.T. Walley, Dover Publications, New York.
\item McKinsey, J.C.C., A.C. Sugar and P. Suppes, 1953, 
`Axiomatic foundations of classical particle mechanics', {\em J. Rational Mechanics and Analysis}, {\bf 2} 253-272.
\item{Sant'Anna, A. S., 1995, `Set-theoretical structure for Hertz's mechanics', in {\em Volume of Abstracts of the 10th International Congress of Logic, Methodology and Philosophy of Science} (LMPS95), International Union of History and Philosophy of Science, 491-491.}
\item Sant'Anna, A.S., 1996, `An axiomatic framework for classical particle mechanics without force', {\em Philosophia Naturalis} {\bf 33} 187-203.
\item Suppes, P., 1957, {\em Introduction to Logic}, Van Nostrand, Princeton.
\item Suppes, P., 1967, {\em Set-Theoretical Structures in Science}, mimeo. Stanford University, Stanford.
\item Suppes, P., 1990, `Philosophy and the sciences', in W. Sieg (editor) {\em Acting and Reflecting}, Kluwer Academic, Dordrecht, 3-30.
\item Tati, T., 1964, `Concepts of space-time in
physical theories', {\em Prog. Theor. Phys. Suppl.} {\bf 29} 1-96.
\item Tati, T., 1983, `The theory of finite degree of
freedom', {\em Prog. Theor. Phys. Suppl.} {\bf 76} 186-223.
\item Tati, T., 1986, `Macroscopic world and
microscopic world: a physical picture of space-time', {\em Annals of the Japan Association for Philosophy of Science} $\bf{7}$ 15-31.
\item Tati, T., 1987, `Local quantum mechanics'', {\em Prog. Theor. Phys.} $\bf{78}$ 996-1008.
\end{enumerate}

\end{document}